\documentclass[journal]{IEEEtran}

\usepackage{graphicx}
\usepackage{subfigure}
\usepackage{float}
\usepackage{amsmath}
\usepackage{amsfonts,amssymb}
\usepackage{dsfont}
\usepackage{algpseudocode}
\usepackage{makecell}
\usepackage{cite}
\usepackage{multirow}
\usepackage{multicol}
\usepackage{booktabs}
\usepackage[ruled,vlined]{algorithm2e}
\usepackage{xcolor,soul}
\usepackage{comment}

\newcommand{\esubsubsection}[1]{\subsubsection{\textit{#1}}\quad}

\usepackage{fancyhdr,lipsum}

\fancypagestyle{mahmood}{%
   \fancyhf{} 
   
   \fancyhead[C]{You can use this material personally. Reprinting or republishing this material for the purpose of advertising or promotion, creating new collective works, reselling or redistributing to servers or lists, or using any copyrighted component in other works must adhere to IEEE policy. The DOI will be supplied as soon as it becomes available.}
}%

\makeatletter
\let\ps@IEEEtitlepagestyle\ps@mahmood

\renewcommand\subsubsection{\@startsection{subsubsection}
{3}
{\z@}
{-1ex\@plus -1ex \@minus -.2ex}
{1ex}{\normalfont\normalsize}}

\begin{document}
\title{An Autonomous Network Orchestration Framework Integrating Large Language Models with Continual Reinforcement Learning}

\author{
    \IEEEauthorblockN{
        Masoud Shokrnezhad\textsuperscript{1}, and Tarik Taleb\textsuperscript{2}
    }
    \IEEEauthorblockA{
       \\
        \textsuperscript{1} \textit{Oulu University, Oulu, Finland}; masoud.shokrnezhad@oulu.fi \\
        \textsuperscript{2} \textit{Ruhr University Bochum, Bochum, Germany}; tarik.taleb@rub.de \\
        \vspace{-20pt}
    }
}




\maketitle

\begin{abstract}
6G networks aim to achieve global coverage, massive connectivity, and ultra-stringent requirements. Space-Air-Ground Integrated Networks (SAGINs) and Semantic Communication (SemCom) are essential for realizing these goals, yet they introduce considerable complexity in resource orchestration. Drawing inspiration from research in robotics, a viable solution to manage this complexity is the application of Large Language Models (LLMs). Although the use of LLMs in network orchestration has recently gained attention, existing solutions have not sufficiently addressed LLM hallucinations or their adaptation to network dynamics. To address this gap, this paper proposes a framework called Autonomous Reinforcement Coordination (ARC) for a SemCom-enabled SAGIN. This framework employs an LLM-based Retrieval-Augmented Generator (RAG) monitors services, users, and resources and processes the collected data, while a Hierarchical Action Planner (HAP) orchestrates resources. ARC decomposes orchestration into two tiers, utilizing LLMs for high-level planning and Reinforcement Learning (RL) agents for low-level decision-making, in alignment with the Mixture of Experts (MoE) concept. The LLMs utilize Chain-of-Thought (CoT) reasoning for few-shot learning, empowered by contrastive learning, while the RL agents employ replay buffer management for continual learning, thereby achieving efficiency, accuracy, and adaptability. Simulations are provided to demonstrate the effectiveness of ARC, along with a comprehensive discussion on potential future research directions to enhance and upgrade ARC.
\end{abstract}

\begin{IEEEkeywords}
6G, Space-Air-Ground Integrated Network (SAGIN), Semantic Communication (SemCom), Network Orchestration, Large Language Model (LLM), Retrieval-Augmented Generation (RAG), Chain-of-Thought (CoT), Reinforcement Learning (RL), Mixture of Experts (MoE), Few-Shot Learning, Contrastive Learning, and Continual Learning.
\end{IEEEkeywords}

\IEEEpeerreviewmaketitle

\section{Introduction}\label{s_intro}
The requirements for sixth-generation (6G) networks include global coverage, which necessitates integrating diverse communication modalities through Space-Air-Ground Integrated Networks (SAGINs). This integration ensures seamless connectivity and consistent service quality in urban, rural, and remote areas \cite{javaid2024leveraging}. Additionally, 6G must enable massive connectivity, ultra-high data rates, and ultra-low latency. Semantic Communication (SemCom) addresses these needs by focusing on message meaning and context rather than bit-oriented transmission, optimizing bandwidth usage and enhancing resource efficiency \cite{shokrnezhad2024semantic}. The problem of orchestrating 6G networks, involving the allocation of resources to users while considering quality of service requirements and system-level objectives, presents inherent complexity. The integration of domains such as space within SAGIN further exacerbates this complexity, as the dynamics of resources remain largely unknown. Furthermore, the implementation of SemCom renders the evaluation of service quality using conventional metrics unfeasible, thereby complicating the problem further \cite{shokrnezhad2024semantic}.

Generally, one potential solution to address decision-making challenges in complex problems where the solution space is excessively large, non-numeric, or characterized by unknown dimensions—rendering algorithmic exploration impractical—is the utilization of Large Language Models (LLMs). This concept has been widely utilized in robotics research, illustrating that LLMs can greatly improve decision-making processes when specific objectives need to be met by identifying an optimal sequence from an infinite array of actions. \cite{pmlr-v205-ichter23a}. However, the use of LLMs in network orchestration remains at an early stage. For instance, Qiu \textit{et al.} \cite{10681550} focused on optimizing the number and placement of wireless access points with LLMs. Sun \textit{et al.} \cite{10778660} adopted LLMs for trajectory planning and wireless resource allocation in scenarios serving Unmanned Aerial Vehicles (UAVs). Liu \textit{et al.} \cite{10520918} equipped Reconfigurable Intelligent Surfaces (RISs) with LLMs to deduce optimized strategies for wireless resource allocation and signal decoding order. Mekrache \textit{et al.} \cite{mekrache2024intent} and Rong \textit{et al.} \cite{rong2024leveraging} expanded this idea to multi-domain resource allocation, addressing network and cloud as well as terrestrial and non-terrestrial integrated infrastructures, respectively. Despite the high quality claimed by these studies, all aspects of resource allocation are exclusively determined by LLMs in their methods, increasing the risk of hallucinations, particularly in expansive solution spaces. Additionally, the adaptation of these LLM-based mechanisms to probable fluctuations, especially unknown ones, has not been investigated. Furthermore, SemCom is entirely overlooked. The only solution superficially proposed is fine-tuning; which is impractical in live setups due to its resource- and time-intensive nature. Thus, the development of a sophisticated LLM-based solution for 6G network orchestration is essential \cite{bariah2024large, maatouk2024large}.

To address the gaps in existing research, we propose a framework named Autonomous Reinforcement Coordination (ARC) for a system of SemCom-enabled SAGIN. To decompose the inherent complexity of the problem, the resource allocation process is structured within a two-tier component called the Hierarchical Action Planner (HAP). In the outer tier, users and their corresponding actions (required resource allocation decisions) are sequenced, while in the inner tier, each action is executed in isolation, following the Mixture of Experts (MoE) concept. The integration of HAP with the system is facilitated by a Retrieval-Augmented Generator (RAG), which gathers data from system elements to prepare inputs for HAP. We utilize LLMs in both RAG, to comprehend strategical textual commands and to evaluate the quality of SemCom connections, and HAP, to derive efficient sequences in the outer tier. For the inner tier of HAP, we utilize pre-trained Reinforcement Learning (RL) agents to minimize the computation and storage overheads associated with training, with each agent assigned to a specific task. While LLMs are reserved for high-level tasks, we implement Chain-of-Thought (CoT) reasoning through few-shot learning to reduce the likelihood of hallucinations to enhance accuracy. To adapt ARC to system dynamics, RAG calculates rewards for recent actions and incorporates them into the inputs of HAP. These rewards are utilized through contrastive learning to select the highest-reward exemplars in few-shot learning, producing updated sequences, and through continual learning to keep the RL agents updated via replay buffer management.

To the best of our knowledge, this is the first work to employ LLMs and related techniques, such as few-shot learning, to enhance continual RL-based resource allocation for improved efficiency, accuracy, and adaptability. The paper is structured as follows. Section \ref{s_foundation} provides essential background by describing the system model, formulating the resource allocation problem, and outlining the associated challenges. Section \ref{s_solution} details the solutions, including the main components and workflows of ARC. An ablation study is presented in Section \ref{s_evaluation}, followed by potential directions for future work  in Section \ref{s_discussion}. Concluding remarks are drawn in Section \ref{s_conclusion}.

\section{Preliminaries}\label{s_foundation}

\subsection{System Model}\label{ss_system_model}
In this paper, we consider a system following the SAGIN model \cite{javaid2024leveraging}, which includes a bi-dimensional terrestrial network layer augmented by a third dimension provided by two non-terrestrial network layers—airborne and spaceborne—at different altitudes and orbits (hereafter referred to as \textit{the infrastructure}). The ground layer, operating at or near the surface of the Earth, consists of cellular and satellite ground stations, interconnected through network nodes via fiber optic and cable links. The air layer, operating in the stratosphere and lower atmosphere, comprises high-altitude platform stations and unmanned aerial vehicles, supporting both cellular and satellite communication interfaces. The space layer, operating in orbit around the Earth, consists of satellite constellations in various orbits (geostationary, low Earth, medium Earth, and highly elliptical). Furthermore, all nodes across the three layers are equipped with computing resources, collectively forming a distributed cloud that provides extensive computational capabilities throughout the network \cite{shokrnezhad2024towards}. The infrastructure can be leveraged as a multi-layered, integrated communication and computation platform to enable innovative services for 6G and beyond. 

We consider a set of predefined services (hereafter referred to as \textit{the services}), each represented by a directed graph of functional blocks. The blocks receive input from a group of active users (hereafter referred to as \textit{the users}) and process outputs sequentially, as specified in the graph, until the desired outcome is achieved. The services are postulated to be semantic-aware, capable of processing semantic segments (or semantics, for the sake of simplicity) rather than conventional bit-oriented traffic, thus enabling semantic communication within the infrastructure. Given this semantic awareness, service quality evaluation is conducted using Quality of Experience (QoE) metrics defined in the semantic space. For instance, in a holographic meeting service in the Metaverse, users' movements, shapes, and spoken words (semantic input traffic) are transmitted to Machine Learning (ML)-based audio generation and video rendering blocks (service graph), with the temporal description of meeting scenes (semantic output traffic) returned to the users for headset visualization. The output quality can be semantically assessed using metrics such as facial expression accuracy, body movement naturalness, and lip sync precision, ensuring a high-fidelity user experience in the metaverse.

\subsection{Problem Formulation}\label{ss_problem_formulation}
To deliver services to users requesting specific functionalities, the resources of the infrastructure must be allocated to service graphs and semantic traffic to meet QoE requirements, a process referred to as the resource allocation process (hereafter \textit{the process}). The system can be modeled as a Markov Decision Process (MDP), encompassing state, action, transition, and objective spaces, along with a reward function. The state space indexes various quality and quantity metrics for each [\textit{user, service, resource}] triplet, including request information that indicates the service requested by the user and the network node connecting the user to the infrastructure. The action space includes all potential resource allocation decisions, such as path selection or functional block replacement. The transition space represents all possible dynamics governing state transitions, reflecting environmental effects. The objective space outlines system-level objectives guiding action selection, such as minimizing cost or maximizing quality. Finally, the reward function provides feedback by mapping a specific set of actions and the resulting state (post-action application) to a scalar value.

\subsection{Challenges and Complexities}\label{ss_challenges_and_complexities}
Suppose that for each objective defined in the objective space, there exists a corresponding policy that governs the selection of actions within the process. The process can then be optimized for each objective by solving the problem of identifying its optimal policy that maximizes the cumulative reward over time (hereafter referred to as \textit{the problem}). The challenges in finding the optimal policy are as follows:

\esubsubsection{Combinatorial Nature}
In the process as formulated above, the optimality of an action is influenced by its preceding and subsequent actions. Thus, identifying the optimal action at each step requires examining all possible sequences of actions, and determining the overall optimal policy necessitates applying this exhaustive investigation to every possible state. This combinatorial complexity renders the problem NP-hard. Moreover, as the number of users, services, and resources increases, the complexity grows exponentially. Consequently, no polynomial-time algorithm can guarantee the discovery of the optimal policy, rendering optimal allocation effectively infeasible for large-scale systems.

\esubsubsection{QoE-based Evaluation}
The NP-hardness of the problem may be mitigated using conventional yet advanced mathematical techniques, but only when the reward function is calculable, which is not the case in this paper. Here, the reward function requires user feedback as part of its input and returns a scalar value. Since this feedback is defined in the semantic QoE space, which could be a descriptive form, there is no closed-form mathematical solution to compare it to the corresponding QoE requirements. Furthermore, combining the outcome with the metrics related to the system-level objective presents an additional challenge, making it extremely difficult to devise the optimal policy.

\esubsubsection{Unknown Transition Dynamics}
Addressing the mitigation of the complexity of the problem also presents another challenge: unknown transition dynamics. Part of our infrastructure resides in space, where various factors can lead to situations affecting the behavior of resources, following an unpredictable pattern. Solar flares and geomagnetic storms can significantly impact satellite trajectories. The behavior of the ionosphere can vary unpredictably, causing rapid fluctuations in the amplitude and phase of radio signals. Moreover, if we extend the SAGIN model to other planets, following the concept of a space internet, interference from unknown sources could affect the dynamics. Furthermore, we have users who may be mobile, with rapid and unpredictable demand shifts. These sudden, unpredictable changes affect the transition dynamics in ways that are difficult to anticipate. When the transition dynamics are unknown, devising the optimal policy becomes a formidable challenge.

\section{Solution} \label{s_solution}
\begin{figure}[!t]
\centerline{\includegraphics[width=3.5in]{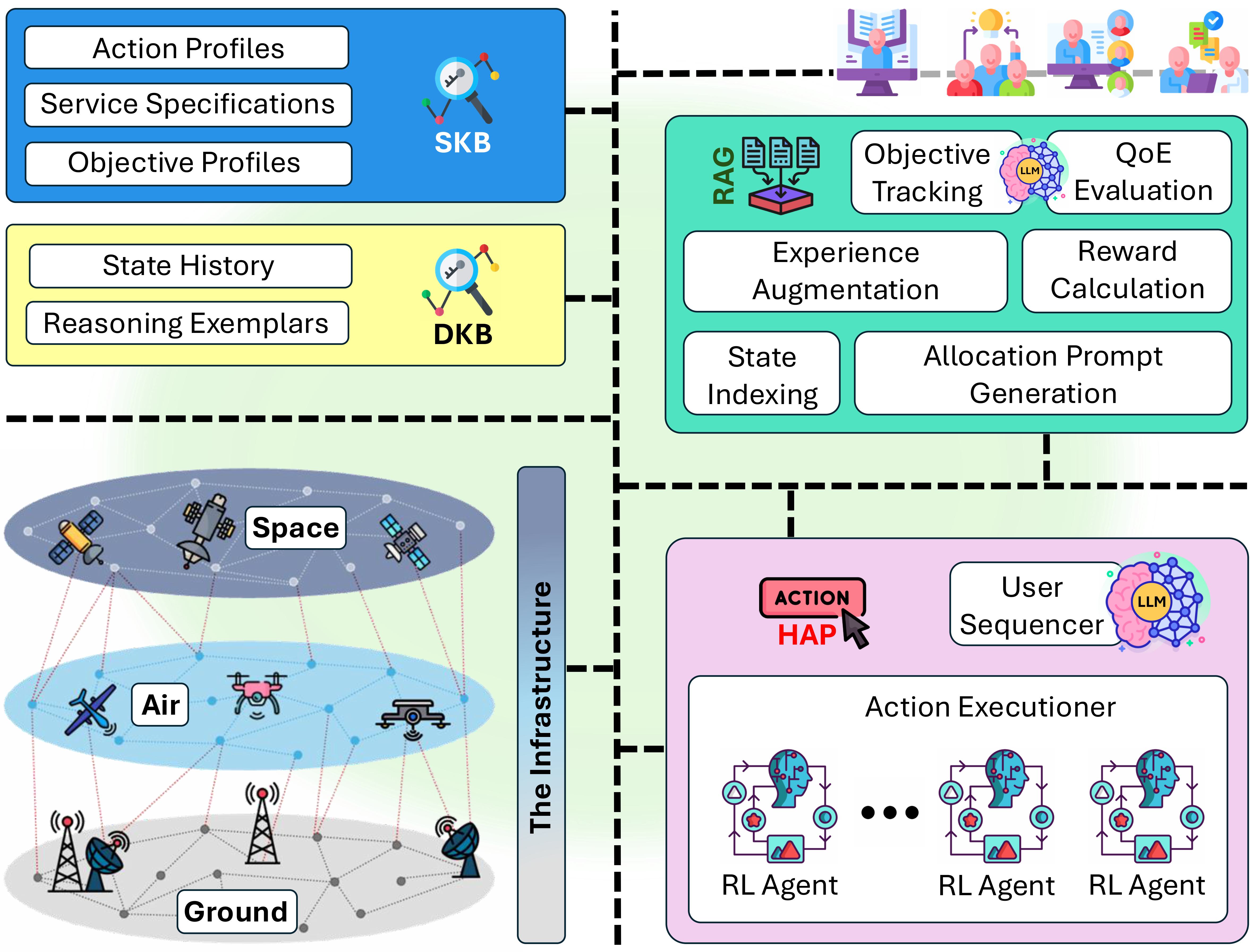}}
\vspace{-5pt}
\caption{The components of ARC.}
\label{fig1}
\vspace{-10pt}
\end{figure}

To address the challenges discussed, we propose a framework called ARC, illustrated in Fig. \ref{fig1}. ARC employs a two-tier hierarchical resource allocation mechanism to effectively manage problem complexity, featuring an LLM-based outer tier for high-level sequencing and a RL-based inner tier for low-level execution of actions. This two-tier orchestrator is integrated into the core of a closed-loop control system through a RAG mechanism, where the state is periodically gathered, resource allocation decisions are made for users as actions, implemented within the infrastructure, and subsequent rewards are assessed and integrated into the process, thereby achieving autonomy. This is detailed in the following two subsections: components and workflows.

\subsection{Components}\label{ss_components}
\esubsubsection{\textbf{S}tatic \textbf{K}nowledge \textbf{B}ase (SKB)}\label{sss_skb}
SKB serves as the foundational database of the system, utilizing a vector-based structure for efficient similarity search of static data. It is designed to encompass the following information:

\textbf{Service Specifications}: Detailed descriptions of the services, with each record corresponding to a service and outlining its directed graph of functional blocks, semantic QoE requirements, and capacity requirements.

\textbf{Objective Profiles}: Three primary objectives, each designed to optimize a specific system-level metric. \textit{Minimizing Cost} prioritizes users with lower capacity requirements and selects resources with the lowest associated costs, where cost is defined as a normalized sum of energy consumption and monetary price. \textit{Maximizing Quality} focuses on users with stringent QoE demands by assigning them to resources that offer superior performance. Lastly, \textit{Load Balancing} ensures an even distribution of resource usage across the infrastructure. At any given time, exactly one of these objectives may be activated and executed in tandem with the primary goal of maximizing the number of supported users. This joint optimization function ensures that a feasible solution is always attainable while effectively managing trade-offs among different performance metrics.

\textbf{Action Profiles}: All possible types of allocation decisions that can be performed over the infrastructure. Concretely, these decisions include choosing the computing resource to host each functional block, allocating the necessary computing and storage capacities, and determining the routing paths through the network. In the space and air layers, this routing involves selecting the next hop, assigning channels, and controlling transmit power to satisfy data rate requirements. In the ground layer, the path is formed by choosing the next hop and allocating the required bandwidth over available links. The overall action space is composed of both discrete elements (e.g., selecting hosts, channels, or next hops from a finite set) and continuous elements (e.g., computing capacity, storage capacity, and power control within allowable ranges).

\esubsubsection{\textbf{D}ynamic \textbf{K}nowledge \textbf{B}ase (DKB)}\label{sss_dkb}
DKB functions as the adaptive repository of the system, and it uses the same vector-based structure to facilitate efficient storage and similarity retrieval. It is structured to incorporate the following dynamic information:

\textbf{State History}: State records of the system, maintained with a window size sufficiently long to capture the patterns of all dynamic behaviors. For instance, starting with a relatively large window assures comprehensive coverage of cyclical patterns, yet the size can be adaptively adjusted—using techniques such as $\epsilon$-greedy—by gradually decreasing it until any performance degradation is observed. 

\textbf{Reasoning Exemplars}: Triples [\textit{input, CoT, output}], where \textit{input} is the combination of a state history and an objective; \textit{CoT} represents a specific sequence of users and actions to support them in the given state history and objective \cite{NEURIPS2022_9d560961}; and \textit{output} is the rewards obtained from applying those actions to the infrastructure. This information is initialized by combining the optimal allocations of solving the problem for different state histories and objectives using mathematical optimization solvers along with random and greedy action selections.

\esubsubsection{\textbf{R}etrieval-\textbf{A}ugmented \textbf{G}enerator (RAG)}\label{sss_rag}
At the heart of ARC lies a component named RAG, which is responsible for coordinating the process of collecting monitoring data from various system elements, processing and enriching it, and subsequently feeding customized inputs to other components. The specific functionalities of this component are detailed as follows. These functions are executed at each time slot.

\textbf{Objective Tracking}: RAG is designed to receive commands from strategists to update the active system-level objective, serving as a means to adjust the system strategy. Upon receiving a command, RAG utilizes an LLM to generate a step-back question to extract its key parts, retrieve its required information, and transform it through expansion. Then, it generates a vector representation and calculates similarity scores (such as maximal cosine) with objective vectors in SKB. The most similar objective is activated. If no command is received, the current objective remains active.

\textbf{QoE Evaluation}: RAG utilizes an LLM to evaluate the resources allocated to users. The LLM calculates a similarity score by comparing user feedback with semantic QoE requirements stored in SKB for their requested services. If the score exceeds a predefined threshold, the requirement is considered met; otherwise, it is not. By integrating the reasoning capabilities of LLMs with structured semantic data, RAG leverages its understanding of multi-modal input to assess alignment and provides nuanced and contextually relevant responses, whereas conventional quality measurement techniques prove inadequate. This ensures that user expectations are effectively aligned with the underlying service quality.

\textbf{State Indexing}: RAG collects data from the infrastructure and the users, extracts the current state in state space format, and encodes it into vector representations for storage in DKB. This process considers QoE evaluation results. If a user's QoE requirements are met, the allocated resources of that user are reflected in the state; otherwise, resources are discarded, and the user is marked as actively seeking resources by adjusting the request info index in the state.

\textbf{Reward Calculation}: Considering the previous state, the active objective, the current state (including QoE evaluation results), RAG calculates the reward for the last actions. If a user's QoE requirements are unmet, the reward for the actions of that user is 0; otherwise, the reward reflects how the allocated resources contribute to optimizing the active objective. RAG publishes this information in the form of a specific prompt, named the update prompt.

\textbf{Experience Augmentation}: RAG updates the reasoning exemplars in DKB, where \textit{input} includes the state history ending with the state of the previous time slot and the active objective associated with it, \textit{CoT} consists of an ordered list of users and actions taken in the previous time slot along with their corresponding rewards, and \textit{output} is the overall reward, which is the sum of the action rewards.

\textbf{Allocation Prompt Generation}: RAG uses an LLM to publish a prompt, called the allocation prompt, indicating that the resources of the infrastructure need to be allocated to a set of users (new users or those with unmet requirements from the previous time slot). This process considers the current state history, the active objective, and the action profiles (stored in SKB) required concerning the specifications of requested services by the users. RAG aslo retrieves reasoning exemplars from DKB with the highest similarity to the current history-objective pair and attaches them to the prompt.

\begin{figure*}[!t]
\centerline{\includegraphics[width=7.1in]{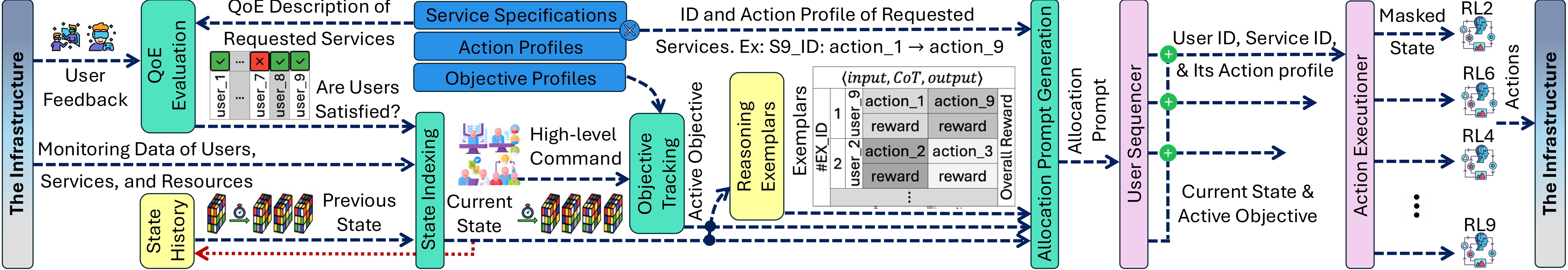}}
\vspace{-5pt}
\caption{The resource allocation workflow in ARC. Note that the color of each sub-component and functionality follows its parent component in Fig. \ref{fig1}.}
\label{fig2}
\end{figure*}

\esubsubsection{\textbf{H}ierarchical \textbf{A}ction \textbf{P}lanner (HAP)}\label{sss_hap}
Knowing the current state history, the active objective, and the action profiles of the users, HAP is the component responsible for actual resource allocation. It interfaces with RAG on one end and connects to the infrastructure resources on the other end. HAP comprises two main sub-components, as follows:

\textbf{User Sequencer}: Our approach to attain optimal allocation in each time slot is to find the optimal sequence of the users such that allocating resources to each user in that sequence in a greedy manner results in an optimal outcome. To show that such a sequence exists, consider the optimal allocation for users in the current time slot, where each user is paired with a specific resource to maximize the total reward. To construct the optimal sequence, start with any user allocated the highest-reward resource in the optimal allocation. Continue selecting users in order of decreasing reward of their allocated resources in the optimal allocation. Now, if we follow this sequence and apply the greedy strategy (always choosing the highest-reward available resource), we will achieve the optimal allocation. Thus, an optimal sequence exists. This sub-component utilizes the inherent comprehension capability of LLMs to use the current state history and the active objective and derive such a sequence for the users. Note that we need to produce the sequence of requests in one shot and not user by user, as LLMs are more adept at generating entire sequences than intermittently pausing and re-initiating the token sampling process.

\textbf{Action Executioner}: This sub-component is responsible for executing action selection for each user, considering the action profiles. To accomplish this, a set of RL agents is employed, with each agent specializing in a specific action. For each user-required action, a masked version of the current state history is generated to restrict the decision-making process to the feasible solution space of the user and the specified service, as well as to the active objective, and is then passed to the respective agent. The resulting decision is subsequently applied to the infrastructure. Decisions regarding the actions of different users should be made in the order of the sequence (after each user, the state history should be updated with its allocation to be used for the next user). However, the strategy for determining the actions of a single user can be selected based on their action profile, allowing for either parallel or serial execution when one action necessitates the outputs of preceding actions. 

\subsection{Workflows}\label{ss_workflows}

\begin{figure*}[!t]
\centerline{\includegraphics[width=7.1in]{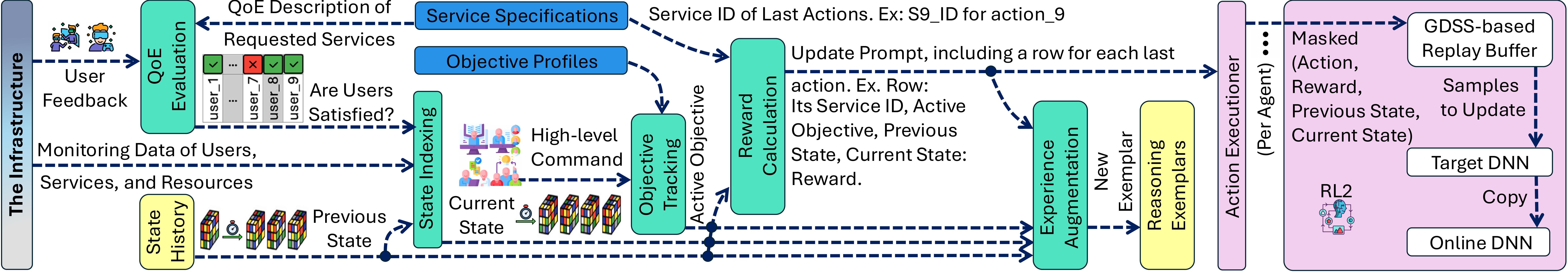}}
\vspace{-5pt}
\caption{The process of updating rewards in ARC to facilitate reward-based few-shot learning and gradient-based continual learning.}
\label{fig3}
\vspace{-10pt}
\end{figure*}

\esubsubsection{Resource Allocation}
The main workflow of ARC is the process of resource allocation, describing how the defined components coordinate to allocate the resources of the users to provision their services. At the begining of the process, illustrated in Fig. \ref{fig2}, the infrastructure is monitored, and QoE evaluation is performed for the users to discover those with broken QoE, shaping the current state history and updating DKB. At the same time, the proper objective is activated. Then, the allocation prompt is created and published. Next, HAP receives this prompt and extracts the users requesting a service from the last state of the state history. The optimal sequence is predicted, and the required actions of each user are executed by the corresponding RL agents one by one. The final step is to apply the allocated resources to the infrastructure. This loop will persist in each time slot. Consequently, the policy defined in Section \ref{ss_challenges_and_complexities} can be redefined as a combination of two HAP-level policies: the sequencer policy and the execution policy (as a function of the policies for each RL agent).

The design of ARC is highly efficient and adaptable, effectively managing inherent complexity while addressing computational overheads by leveraging the capabilities of LLMs. Transforming raw information into a coherent sequence of actions is accomplished by an LLM, which demonstrates its ability to efficiently synthesize complex data without introducing excessive computational burdens. Such mechanim also promotes the integration of various systems with different types of resources, enhancing practicality in aligning with existing network architectures. To execute the extracted actions, a MoE approach is utilized, where each action is managed by a dedicated RL agent. This specialization simplifies the agents, making them easier to debug and upgrade while facilitating real-time feasibility in decision-making processes. Such a design facilitates the scalability of ARC; the exponential growth in LLM capabilities allows for the expansion of infrastructure, resources, and services. New objectives and actions can also be introduced, as they can be managed by separate agents without altering existing ones. Additionally, the incorporation of reward-enabled contrastive and continual learning further enhances the adaptability of ARC to changes in the environment.

\esubsubsection{Lifelong Adaptation}

To optimize the resource allocation and maintain its efficiency regarding the combinatorial nature of the problem while addressing unknown transmission dynamics, the strategy focuses on enhancing the sequencer and execution policies through mechanisms of ARC, as detailed below.

\textbf{Reward-based Few-Shot Learning}: To improve the sequencer policy, we employ few-shot learning. Using an LLM as the user sequencer, few-shot learning involves adding a few reasoning exemplars to the allocation prompt. As shown in Fig. \ref{fig2} and described in Section \ref{sss_dkb}, the reasoning exemplar includes a chain of thoughts demonstrating the sequence of users and their corresponding actions selected for a specific history-objective pair. Leveraging the vast training data by LLMs, which provides a deep understanding of reasoning patterns, they can utilize the reasoning chain in the exemplars to infer the desired reasoning process and gravitate towards prioritizing appropriate sequences for given state histories and active objectives. Moreover, these exemplars serve to reduce the likelihood of hallucinations by presenting the model with numerous valid responses, thereby guiding it toward more accurate outputs. Simultaneously, they increase the probability of exploring the solution space effectively, as the model is exposed to a diverse range of acceptable answers. To further enhance the efficiency of few-shot learning, we employ contrastive learning, where RAG selects two groups of reasoning exemplars for the allocation prompt: one with the highest rewards and one with the lowest. This approach provides the LLM with updated instructions on what to pursue and what to avoid, helping it remain adapted to changes. This is because reasoning exemplars are updated over time in DKB, and those reflecting patterns valid for the current transition dynamics are expected to yield higher rewards.

\textbf{Gradient-Based Continual Learning}: 
To enhance the execution policy, we utilize Continual Reinforcement Learning (CRL) with agents structured around two online and target Deep Neural Networks (DNNs) and a replay buffer \cite{mazandarani2024semantic}. Model updates occur through batch training using the target DNN and data from the replay buffer, which is refreshed at each time slot via the update prompt, as illustrated in Fig. \ref{fig3}. Action selection is performed by the online DNN, which is replaced by the target DNN over a longer period, allowing agents to adapt to system changes while maintaining stability. To prevent catastrophic forgetting, we implement Gradient-Based Sample Selection (GDSS) \cite{NEURIPS2019_e562cd9c}, which effectively stores training data in the replay buffer to represent the agent's entire history. This method assigns scores to new samples based on cosine similarity with existing samples and replaces less effective ones, maximizing diversity in the replay buffer. Alongside the modular design, which simplifies the retraining processes, we further reduce the computation and storage overheads of retraining with CRL while enabling agents to retain past transition dynamics.

\section{Evaluation} \label{s_evaluation}

\begin{figure}[!t]
\centerline{\includegraphics[width=3.3in]{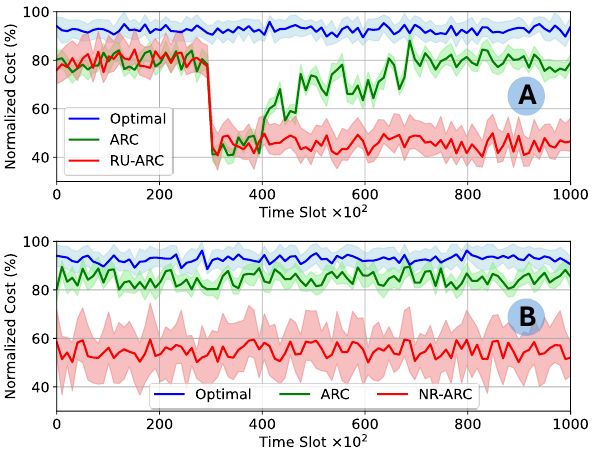}}
\vspace{-5pt}
\caption{Two scenarios comparing the results of ARC with optimal, A) reward-unaware, and B) non-reinforcement results. The normalized cost of each allocation is calculated to ensure that allocations with lower costs receive higher scores, and vice versa.}
\label{fig4}
\vspace{-10pt}
\end{figure}

In this section, we evaluate ARC through numerical simulations using an infrastructure of 10 nodes, half of which are non-terrestrial and follow a predefined movement pattern to maintain periodic connectivity and connect to three nearest neighbors. Each node is equipped with computing resources ranging from 10 to 100 MIPS (Million Instructions Per Second). The characteristics of the links between these nodes vary based on distance, with link latency ranging from 1 to 10 ms (milliseconds) and link capacities fluctuating between 10 to 100 Mbps (Megabits per second); the associated costs have the same numerical values as the link capacities. We analyze a set of 10 users requesting services involving a single functional block, whose computing capacity is uniformly selected from 2 to 5 MIPS. The service processes images with a QoE requirement that states: \textit{the image must be 1920x1080 to be acceptable}, with dimensions depending on the allocated network link capacity. Users' action profiles involve placing the block on a node and establishing connections while meeting their requirements. Our goal is to minimize allocation costs, with rewards calculated by dividing the average resource cost by each action's allocated cost, thereby favoring low-cost allocations. To initialize exemplars, we use mathematical optimization solvers to determine optimal solutions at each time slot, as described in Section \ref{sss_hap}. For simulations, we employ LLaMA 3.1-8B without fine-tuning, implementing the Double Dueling Deep Q-Learning (D3QL) technique for reinforcement learning agents \cite{mazandarani2024semantic}. 

To evaluate ARC, we conduct an ablation study in two scenarios: first, assessing the impact of incorporating rewards, and second, examining the effect of integrating RL-based low-level action execution. The results are shown in Fig. \ref{fig4}.A and Fig. \ref{fig4}.B. In the first scenario, we compare ARC with Reward-Unaware ARC (RU-ARC), which disregards reward calculations and provides HAP with reasoning exemplars based solely on similarity, without training RL agents, which operate on pre-trained weights. The findings indicate that ARC allocates resources efficiently, achieving near-optimal cost outcomes. After a transition at iteration 30,000, during which we modified the link latency calculation by swapping low-latency links with high-latency ones to invalidate the previous allocations and compel the framework to select new actions, ARC exhibits a notable recovery capability, whereas RU-ARC fails to demonstrate similar resilience. In the second scenario, we implement Non-Reinforcement ARC (NR-ARC), eliminating RL agents and allowing the LLM to manage low-level allocations. Results show NR-ARC performs worse than ARC, displaying greater fluctuations due to LLM instability. These findings demonstrate that the components of ARC work together effectively, enabling it to extract efficient action sequences under stable conditions while also swiftly adapting to dynamic challenges. This synergy underscores ARC's efficiency, accuracy, and adaptability, especially when compared to prior approaches that employed LLMs for resource allocation decision-making without incorporating feedback mechanisms. Such capabilities are crucial for the varied demands of 6G applications, reinforcing ARC's suitability for a wide range of use cases within 6G.

\section{Discussion}\label{s_discussion}

\textbf{Prediction-based State Indexing}: 
Explicitly predicting future states and concatenating them with state history is beneficial for preventing service disruptions, especially in fluctuating environments where proactive decision-making is essential. To achieve this, a Generative Adversarial Network (GAN)-based solution can be employed, using a Long Short-Term Memory (LSTM) network as the generator to capture temporal dependencies from historical data, while a Recurrent Neural Network (RNN) serves as the discriminator to evaluate the predicted state.

\textbf{Autonomous Service/Action Extension}: 
By expanding the coverage of the system into unfamiliar territories, it is expected that users will request unknown services with new QoE requirements, thereby necessitating the creation of unknown action profiles across new resources. One potential solution is to utilize an LLM-based approach to define the unknown service as a combination of existing services. A similar methodology can be applied to establish new action profiles based on the existing ones.

\textbf{Online LLM Training}: 
Research indicates that few-shot learning significantly enhances the reasoning capabilities of LLMs, sometimes making them competitive with costly fine-tuning approaches in terms of time and resources. However, the efficiency of the user sequencer relies on the quality of allocation prompts generated by RAG, which does not impact fine-tuned LLMs. To address this dependency, a potential solution is to use an online LLM for inference alongside a target LLM trained with data collected from the infrastructure.


\textbf{Towards Algorithm-of-Thoughts (AoT)}: 
A fundamental aspect of our proposed approach is the initial records of reasoning exemplars, generated by solving the resource allocation problem with mathematical solvers. Since the exact step-by-step process for resource allocation is unknown, this may result in unpredictability. When optimality can be sacrificed for determinism, we can substitute CoT reasoning with an Algorithm-of-Thoughts (AoT)-based approach \cite{sel2023algorithm}. This involves designing a heuristic algorithm tailored to each objective and defining it in each exemplar with specific inputs, enabling the LLM to internalize the algorithm and generate responses that mimic algorithmic search.

\section{Conclusion}\label{s_conclusion}
In this paper, we propose a framework named ARC for a SemCom-enabled SAGIN, which utilizes a hierarchical action planner supported by a retrieval-augmented generator to orchestrate network resources. ARC decomposes orchestration into two tiers, employing an LLM for high-level planning and RL agents for low-level action execution. The LLM utilizes CoT reasoning for few-shot learning, augmented by contrastive learning, while the RL agents implement replay buffer management for continual learning, thereby achieving efficiency, accuracy, and adaptability. Following the demonstration of the performance of ARC through proof-of-concept simulations, we outline detailed future directions to enhance the performance of ARC and expand its capabilities, including prediction-based state indexing, autonomous service/action extension, online LLM training, one-shot inference, and integration with AoT.


\bibliographystyle{IEEEtran}
\bibliography{main}

\vspace{10pt}
\footnotesize
\noindent \textbf{Masoud Shokrnezhad}, Ph.D., is a postdoctoral researcher at University of Oulu, Finland, focusing on semantic-aware resource orchestration for 6G.

\vspace{5pt}
\noindent \textbf{Tarik Taleb} is a Full Professor at Ruhr University Bochum, Germany, the founder of ICTFicial Oy, and the director of the MOSA!C Lab in Espoo, Finland, specializing in  softwarization for beyond 5G and 6G networks.

\end{document}